\documentclass[prl,aps,twocolumn,superscriptaddress,showpacs]{revtex4}
\usepackage{amsmath,amssymb,graphicx}
\usepackage{pxfonts}
\usepackage{graphicx}
\usepackage{color}
\usepackage{float}
\begin{document}

\date{July 1, 2023}

\title{Transport and localization of indirect excitons in a van der Waals heterostructure}

\author{L.~H.~Fowler-Gerace$^*$}
\affiliation{Department of Physics, University of California at San Diego, La Jolla, CA 92093, USA}
\author{Zhiwen~Zhou$^*$}
\affiliation{Department of Physics, University of California at San Diego, La Jolla, CA 92093, USA}
\author{E.~A.~Szwed}
\affiliation{Department of Physics, University of California at San Diego, La Jolla, CA 92093, USA}
\author{D.J. Choksy}
\affiliation{Department of Physics, University of California at San Diego, La Jolla, CA 92093, USA}
\author{L.~V.~Butov} 
\affiliation{Department of Physics, University of California at San Diego, La Jolla, CA 92093, USA}

\begin{abstract}
\noindent
Long lifetimes of spatially indirect excitons (IXs), also known as interlayer excitons, allow implementing both quantum exciton systems and long-range exciton transport. Van der Waals heterostructures (HS) composed of atomically thin layers of transition-metal dichalcogenides (TMD) offer the opportunity to explore IXs in moir{\'e} superlattices. The moir{\'e} IXs in TMD HS form the materials platform for exploring the Bose-Hubbard physics and superfluid and insulating phases in periodic potentials. IX transport in TMD HS was intensively studied and diffusive IX transport with $1/e$ decay distances $d_{1/e}$ up to $\sim 3$~$\mu$m was realized. In this work, we present in MoSe$_2$/WSe$_2$ HS the IX long-range transport with $d_{1/e}$ exceeding 100~$\mu$m and diverging at the optical excitation resonant to spatially direct excitons. 
The IX long-range transport vanishes at high temperatures.
With increasing IX density, IX localization, then IX long-range transport, and then IX reentrant localization is observed. 
The results are in qualitative agreement with the Bose-Hubbard theory of bosons in periodic potentials predicting superfluid at $N \sim 1/2$ and insulating at $N \sim 0$ and $N \sim 1$ phases for the number of bosons per site of the periodic potential $N$.
\end{abstract}

\maketitle

A spatially indirect exciton (IX) is a bound pair of an electron and a hole confined in separated layers~\cite{Lozovik1976}. Due to the spatial separation of electrons and holes, the lifetimes of IXs can exceed the lifetimes of spatially direct excitons (DXs) by orders of magnitude. The long lifetimes allow IXs to travel long distances. The long-range IX transport has been extensively studied in GaAs HS where the $1/e$ decay distances of IX luminescence $d_{1/e}$ reach tens and hundreds of microns~\cite{Hagn1995, Larionov2000, Butov2002, Voros2005, Ivanov2006, Gartner2006, High2008, Remeika2009, Vogele2009, Lazic2010, High2012, Alloing2012, High2013, Lazic2014, Leonard2018, Leonard2021}. The long-range IX transport in GaAs HS enabled the observation of various exciton transport phenomena including the exciton ring~\cite{Ivanov2006}, the excitonic transistor~\cite{High2008}, the exciton delocalization~\cite{Remeika2009}, and the coherent exciton transport~\cite{High2012, Leonard2021} and spin transport~\cite{High2013, Leonard2018}.

The temperature of quantum degeneracy, which can be achieved for excitons, scales proportionally to the exciton binding energy $E_{\rm X}$~\cite{Fogler2014}. IXs in GaAs HS have low $E_{\rm X} \lesssim 10$~meV~\cite{Zrenner1992, Sivalertporn2012}. For IXs in III-V and II-VI semiconductor HS, including GaAs~\cite{Hagn1995, Larionov2000, Butov2002, Voros2005, Ivanov2006, Gartner2006, High2008, Remeika2009, Vogele2009, Lazic2010, High2012, Alloing2012, High2013, Lazic2014, Leonard2018, Leonard2021}, GaN~\cite{Chiaruttini2019}, and ZnO~\cite{Morhain2005} HS, the highest $E_{\rm X}\sim 30$~meV is in ZnO HS.

Excitons with remarkably high binding energies can be realized in van der Waals HS composed of atomically thin layers of TMD~\cite{Geim2013, Ye2014, Chernikov2014, Goryca2019}. The IX binding energies in TMD HS reach hundreds of meV~\cite{Deilmann2018}, making TMD HS the materials platform for the realization of high-temperature IX quantum phenomena~\cite{Fogler2014, Berman2017}.

TMD HS offer the opportunity to explore IXs in moir{\'e} superlattice potentials with the period $b \approx a/{\sqrt{\delta \theta^2 + \delta^2}}$ typically in the $\sim 10$~nm range ($a$ is the lattice constant, $\delta$ the lattice mismatch, $\delta \theta$ the deviation of the twist angle between the layers from $i\pi/3$, $i$ is an integer)~\cite{Wu2018, Yu2018, Wu2017, Yu2017, Zhang2017a, Rivera2018, Seyler2019, Tran2019, Jin2019, Alexeev2019, Shimazaki2020, Wilson2021, Gu2022}. IXs are out-of-plane dipoles and the interaction between IXs is repulsive~\cite{Yoshioka1990}. IXs in moir{\'e} superlattices provide the experimental realization of the 2D Bose-Hubbard model for bosons with repulsive dipolar interaction. 

The moir{\'e} potentials can be affected by atomic reconstruction~\cite{Weston2020, Rosenberger2020} and by disorder. The realization of fairly periodic moir{\'e} potentials, in particular on long scales, requires that the disorder is weak. 

Transport of both DXs in TMD monolayers (MLs)~\cite{Kumar2014, Kulig2018, Cadiz2018, Leon2018, Leon2019, Hao2020, Datta2022} and IXs in TMD HS~\cite{Rivera2016, Jauregui2019, Unuchek2019a, Unuchek2019, Liu2019, Choi2020, Huang2020, Yuan2020, Li2021, Wang2021, Shanks2022, Sun2022, Tagarelli2023} was intensively studied. Diffusive IX transport with $1/e$ decay distances $d_{1/e}$ up to a few~$\mu$m was realized in these studies. Disordered potentials can be responsible for limiting the IX transport distances in TMD HS. IX transport in TMD HS can be facilitated by voltage-induced suppression of moir{\'e} potentials~\cite{Yu2017, Fowler-Gerace2021} or by moving IXs with acoustic waves~\cite{Peng2022}.

In this work, we present in MoSe$_2$/WSe$_2$ HS the IX long-range transport with $d_{1/e}$ exceeding 100~$\mu$m and diverging at the optical excitation resonant to DXs. 
The IX long-range transport vanishes at high temperatures.
With increasing IX density, IX localization, then IX long-range transport, and then IX reentrant localization is observed. 
The results are in qualitative agreement with the Bose-Hubbard theory of bosons in periodic potentials predicting superfluid at $N \sim 1/2$ and insulating at $N \sim 0$ and $N \sim 1$ phases for the number of bosons per site of the periodic potential $N$~\cite{Fisher1989}.

We study MoSe$_2$/WSe$_2$ HS assembled by stacking mechanically exfoliated 2D crystals. IXs are formed from electrons and holes confined in adjacent MoSe$_2$ and WSe$_2$ MLs, respectively, encapsulated by hBN layers. No voltage is applied in the HS. IXs form the lowest-energy exciton state in the MoSe$_2$/WSe$_2$ HS. The HS details are presented in Supporting Information (SI).

\begin{figure*}
\begin{center}
\includegraphics[width=15cm]{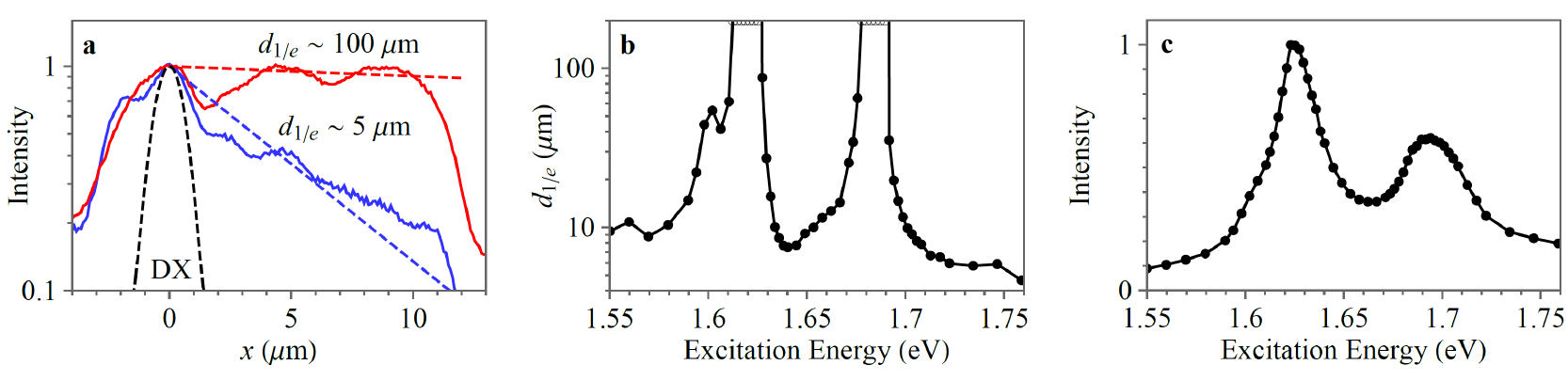}
\caption{The long-range IX transport realized with optical excitation resonant to DXs. (a) Normalized IX luminescence profiles for laser excitation off ($E_{\rm ex} = 1.771$~eV, blue) and near ($E_{\rm ex} = 1.676$~eV, red) the DX absorption resonance. The HS active area extends from $x = -3$ to 11~$\mu$m. The blue and red dashed lines show decays with $d_{1/e} = 5$ and 100~$\mu$m, respectively. The black dashed line shows the DX luminescence profile in the WSe$_2$ ML, this profile is close to the laser excitation profile for short-range DX transport. 
(b) The $1/e$ decay distance of IX luminescence $d_{1/e}$ vs. $E_{\rm ex}$. $d_{1/e}$ are obtained from least-squares fitting the IX luminescence profiles to exponential decays in the region $x = 0 - 11$~$\mu$m. The data with the fit indicating diverging $d_{1/e}$ are presented by points on the edge. 
(c) Integrated IX luminescence intensity vs. $E_{\rm ex}$ showing two absorption peaks corresponding to the MoSe$_2$ and WSe$_2$ DXs. 
$P_{\rm ex} = 0.2$~mW. $T = 1.7$~K.
The $\sim 1.5$~$\mu$m laser spot is centered at $x = 0$. 
}
\end{center}
\label{fig:spectra}
\end{figure*}

The long-range IX propagation with the diverging decay distance $d_{1/e}$, that is with no IX luminescence decay within the entire HS (Fig.~1b), is realized when the optical excitation has the energy $E_{\rm ex}$ close to the MoSe$_2$ or WSe$_2$ DX energy (Fig.~1c). The HS dimensions allow establishing that the longest $d_{1/e}$ exceed 100~$\mu$m. In contrast, for a non-resonant excitation, $d_{1/e}$ is substantially shorter (Fig.~1a,b). 

IXs have built-in electric dipoles $\sim ed_z$ ($d_z$ is the separation between the electron and hole layers) and the interaction between IXs is repulsive~\cite{Yoshioka1990}. An enhancement of IX transport with increasing IX density can be caused (i) by the suppression of IX localization and scattering and (ii) by the IX-interaction-induced drift from the origin. The data outlined in the next paragraph show that the first factor, the suppression of IX localization and scattering, causes the major effect.

The nature of the second factor is an increase of IX energy at the excitation spot $\delta E$ with increasing IX density $n$ (Fig.~2a) that causes IX drift from the origin~\cite{Ivanov2006}. The resonant excitation produces a higher $n$ due to a higher absorption, thus increasing $\delta E$ and, in turn, the IX drift. However, the higher $n$ and $\delta E$ can be also achieved for nonresonant excitation using higher excitation powers $P_{\rm ex}$. Figure 2b shows $d_{1/e}$ vs. $\delta E$ both for the nearly resonant and nonresonant excitation. For the same $\delta E$, a significantly higher $d_{1/e}$ is realized for the (nearly) resonant excitation. This shows that the effect of IX energy increase at the excitation spot on the enhancement of IX transport at resonant excitation is minor and, in turn, that the strong enhancement of IX transport at resonant excitation originates from the suppression of IX localization and scattering. This also shows that the resonant excitation is essential for the realization of the long-range IX transport. 

\begin{figure}
\begin{center}
\includegraphics[width=8.5cm]{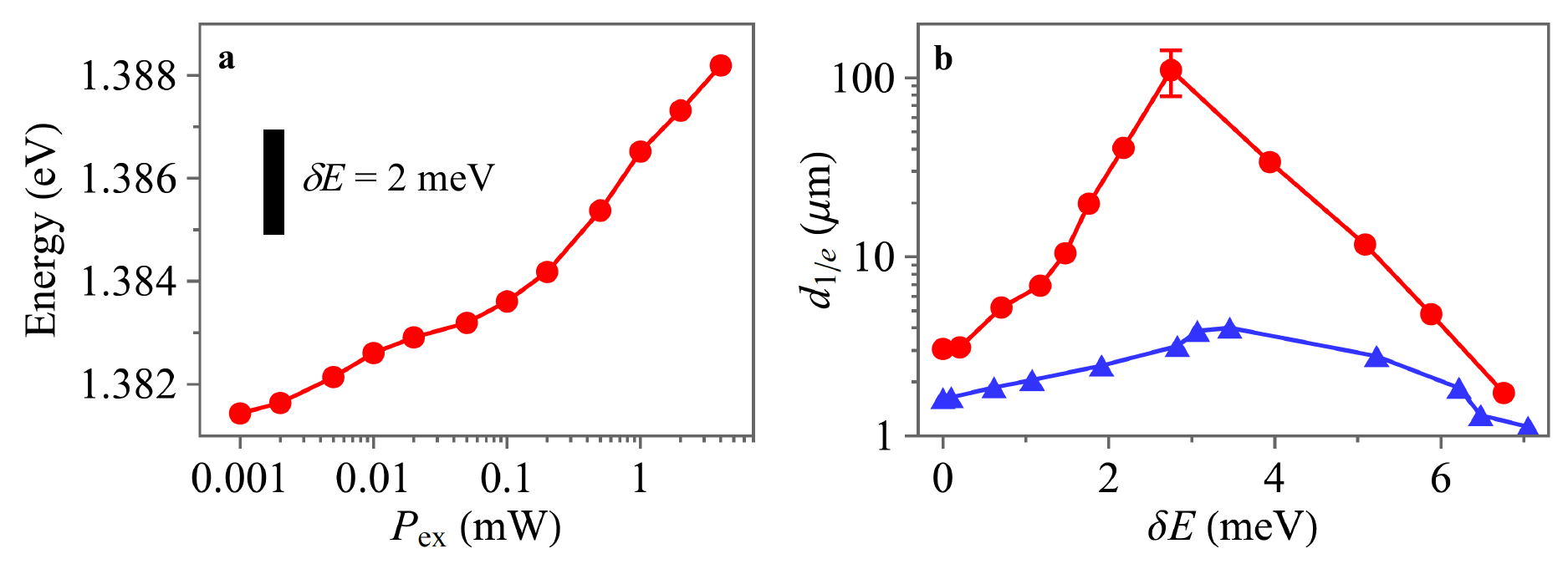}
\caption{The IX transport variation with density at resonant and nonresonant excitation. 
(a) The IX energy at the excitation spot vs. $P_{\rm ex}$. $E_{\rm ex} = 1.623$~eV. The scale bar corresponds to the IX energy increase $\delta E = 2$~meV. 
(b) The $1/e$ decay distance of IX luminescence $d_{1/e}$ vs. $\delta E$ for resonant ($E_{\rm ex} = 1.623$~eV, red) and nonresonant ($E_{\rm ex}$ = 1.96~eV, blue) excitation. 
$T = 6$~K. 
}
\end{center}
\label{fig:spectra}
\end{figure}

The IX energy increase due to the repulsive interaction $\delta E$ (Fig.~2a) can be used for estimating the IX density $n$. For the long-range IX transport observed at $P_{\rm ex} = 0.2$~mW (Figs.~3a, 4a), $\delta E \sim 3$~meV (Fig.~2) and an estimate for $n$ using the 'plate capacitor' formula $\delta E = n u_0$ with $u_0 = 4\pi e^2 d_z / \varepsilon$~\cite{Yoshioka1990} gives $n \sim 2 \times 10^{11}$~cm$^{-2}$ ($d_z \sim 0.6$~nm, the dielectric constant $\varepsilon \sim 7.4$~\cite{Laturia2018}). The estimated temperature of quantum degeneracy $T_{\rm q} = 2\pi \hbar^2n/(m k_{\rm B}) \sim 10$~K~\cite{Fogler2014} ($m \sim 0.9 m_0$ is the IX mass in the TMD HS~\cite{Kormanyos2015}). These estimates indicate that the long-range IX transport is observed at temperatures $\lesssim T_{\rm q}$ and at IX densities below the Mott transition density $n_{\rm Mott} \gtrsim 10^{12}$~cm$^{-2}$~\cite{Fogler2014, Wang2019}. 

Regarding the role of resonant excitation, heating of the IX system by the excitation close to the DX resonances is, in general, smaller than for non-resonant excitation. In particular, the colder IXs created by the resonant excitation screen the disorder more effectively that can facilitate the emergence of IX superfluidity~\cite{Ivanov2006, Remeika2009, Nikonov1998}. 

IXs in moir{\'e} superlattices are repulsively interacting excitons in periodic potentials. For repulsively interacting bosons in periodic potentials, the Bose-Hubbard model predicts both the superfluid phase and insulating phases, such as the Mott insulator and the Bose glass, with the superfluid more stable for the number of particles per lattice site $N \sim 1/2$ and the insulating for $N \sim 1$~\cite{Fisher1989}. The comparison of the data with these predictions shows that the observed IX transport is consistent with the Bose-Hubbard model for repulsively interacting particles in periodic potentials~\cite{Fisher1989}. 
(In contrast, the observed IX transport is inconsistent with the classical drift-diffusion model ~\cite{Ivanov2006} as outlined in SI.) 

The variation of IX transport with temperature and excitation power is presented by the variation of $d_{1/e}$ in Fig.~3a,b. The measured $P_{\rm ex}-T$ diagram for the IX transport distance $d_{1/e}$ is shown in Fig.~3c.

\begin{figure}
\begin{center}
\includegraphics[width=8.3cm]{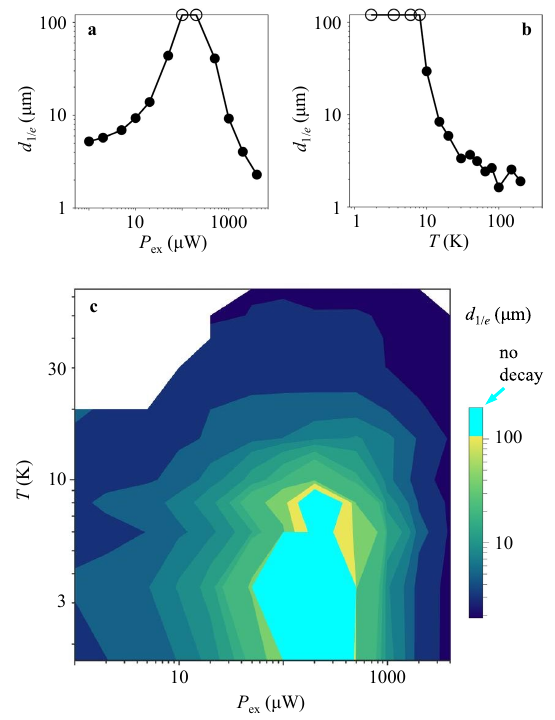}
\caption{Excitation power and temperature dependence of IX transport. 
(a,b) The $1/e$ decay distance of IX luminescence $d_{1/e}$ vs. $P_{\rm ex}$ (a) and vs. temperature (b).  
(c) $d_{1/e}$ vs. $P_{\rm ex}$ and temperature. 
$d_{1/e}$ are obtained from least-squares fitting the IX luminescence profiles to exponential decays in the region $x = 0 - 9$~$\mu$m. The data with the fit indicating diverging $d_{1/e}$ are presented by points on the edge (a,b) or by cyan color (c).
$T = 3.5$ K (a), $P_{\rm ex} = 0.2$ mW (b), $E_{\rm ex} = 1.689$~eV (a-c).
}
\end{center}
\label{fig:spectra}
\end{figure}

For classical IX drift and diffusion, an in-plane potential suppresses IX transport and a long-range IX transport is observed when the IX interaction energy or thermal energy become comparable to the amplitude of in-plane potential~\cite{Ivanov2006, Remeika2009}. However, the predicted amplitude of moir{\'e} superlattice potential is in the range of tens of meV~\cite{Wu2018, Yu2018, Wu2017, Yu2017, Zhang2017a} and is much larger than both the interaction energy $\delta E \sim 3$~meV and thermal energy  $k_{\rm B}T \lesssim 1$~meV for the observed long-range IX transport. This indicates that the nature of the long-range IX transport is beyond classical drift and diffusion. For IX transport over distances much larger than the moir{\'e} superlattice period, the effect of moir{\'e} potential on IX transport is qualitatively similar for R and H (AA and AB) stacking: Both for R and H stacking, IXs propagate in a periodic moir{\'e} potential with the period $b \sim a /\delta \theta$ and large amplitude. The values for the latter for R and H stacking are $E_{\rm R} \sim 100$~meV and $E_{\rm H} \sim 25$~meV~\cite{Wu2018, Yu2018, Wu2017, Yu2017, Zhang2017a}. Since both $E_{\rm R}$ and $E_{\rm H}$ are much larger than $\delta E$ and $k_{\rm B}T$, the discussions of IX transport apply to both R and H stacking. 

The suppression of IX transport with density is observed at high densities (Fig.~3a). This is also inconsistent with classical IX drift and diffusion. For classical IX drift and diffusion, the density dependence is the opposite: both the IX diffusion coefficient and the IX drift increase with density, the former due to the enhanced screening of in-plane potential and the latter due to the enhanced $\delta E$ at the origin~\cite{Ivanov2006, Remeika2009}, as outlined in SI. An enhancement of IX transport with density due to these factors is observed for IXs in both GaAs and TMD HS~\cite{Ivanov2006, Remeika2009, Jauregui2019, Wang2021, Sun2022}. 
In contrast, the reentrant localization of IXs at densities higher than the densities corresponding to the long-range transport of IXs (Fig.~3a) is observed for IXs in TMD HS with periodic moir{\'e} potential. 

\begin{figure*}
\begin{center}
\includegraphics[width=15cm]{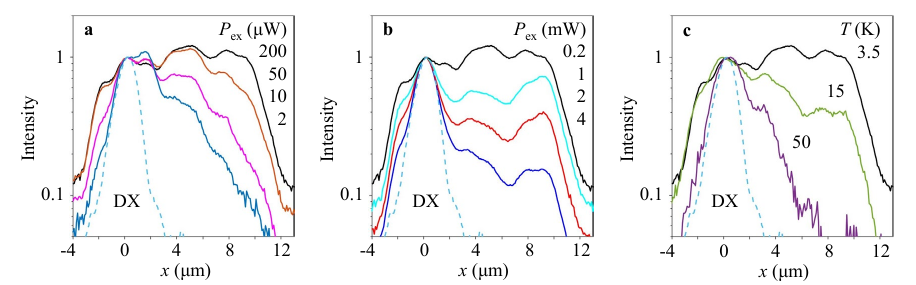}
\caption{Excitation power and temperature dependence of IX transport. 
(a-c) Normalized IX luminescence profiles for different $P_{\rm ex}$ (a,b) and temperatures (c). The dashed line shows the DX luminescence profile in the  MoSe$_2$ ML, this profile is close to the laser excitation profile for short-range DX transport. 
$T = 3.5$~K (a,b), $P_{\rm ex} = 0.2$~mW (c), $E_{\rm ex} = 1.689$~eV (a-c). The $\sim 2$~$\mu$m laser spot is centered at $x = 0$.
}
\end{center}
\label{fig:spectra}
\end{figure*}

The observed enhancement followed by the suppression of IX transport with density (Fig.~3a) is in agreement with the predictions of the Bose-Hubbard model for quantum transport of bosons in periodic potentials. The number of IXs per site of the moir{\'e} superlattice $N$ increases with density. For the maximum IX transport distances (Figs.~2, 3a), the IX density $n$ estimated from the energy shift $\delta E = n u_0$~\cite{Yoshioka1990} as outlined above is $\sim 2 \times 10^{11}$~cm$^{-2}$. $N \sim 1/2$ at $n \sim 2 \times 10^{11}$~cm$^{-2}$ for the superlattice period $b = 17$~nm. This period $b \sim a / \delta \theta$ corresponds to the twist angle $\delta \theta = 1.1^\circ$, which agrees with the angle between MoSe$_2$ and WSe$_2$ edges in the HS (Fig.~S2c), see SI. The IX transport is suppressed for the higher $P_{\rm ex}$ realizing the higher $\delta E$, $n$, and $N$ (Figs.~2, 3a). 
This estimate indicates that the IX long-range transport and localization are in qualitative agreement with the Bose-Hubbard theory of bosons in periodic potentials predicting superfluid at $N \sim 1/2$ and insulating at $N \sim 0$ and $N \sim 1$ phases for the number of bosons per site of the periodic potential $N$.

The IX long-range transport with the diverging decay distance vanishes at $T \sim 10$~K (Fig.~3b). This is in qualitative agreement with the transition to a non-superfluid phase with increasing temperature~\cite{Fisher1989}. (In contrast, for the classical diffusive IX transport~\cite{Ivanov2006}, the temperature dependence is the opposite: the IX diffusion increases with temperature, as outlined in SI.)

The theory predicts that the critical temperature for the superfluid-normal transition for bosons in periodic potentials $T_{\rm c} \sim 4\pi N J$~\cite{Capogrosso-Sansone2008}. The inter-site hopping $J$ is determined by the lattice parameters and drops with increasing lattice amplitude $E_{\rm l}$. $T_{\rm c} \sim 10$~K corresponds to $J \sim 1.6$~K~\cite{Capogrosso-Sansone2008} and, in turn, for the moir{\'e} potential with $b \sim 17$~nm, to $E_{\rm l} \sim 8$~meV~\cite{Remeika2012}. This value is smaller than the amplitudes considered in Refs.~\cite{Wu2018, Yu2018, Wu2017, Yu2017, Zhang2017a}. However, the estimates here are rough and developing the theory of IX superfluidity in moir{\'e} potentials is needed for a more accurate comparison. 

The theory predicts that higher $T_{\rm c}$ can be achieved in lattices with higher $J$~\cite{Capogrosso-Sansone2008}. This can be realized in moir{\'e} superlattices with smaller periods in HS with larger twist angles between the TMD layers.

The theory also predicts that at $N \sim 1$, the Mott insulator form for sufficiently high on-site interaction $U$ ($U \gtrsim 17 J$)~\cite{Capogrosso-Sansone2008}. For the considered moir{\'e} superlattice, the estimated $U \gtrsim 10$~meV exceeds the required value and the insulating phase should form at $N \sim 1$ according to the theory. This is in agreement with IX localization at $N \sim 1$ observed in the experiment. 

In summary, this work presents in a MoSe$_2$/WSe$_2$ HS the IX long-range transport with the decay distance $d_{1/e}$ exceeding 100~$\mu$m and diverging at the optical excitation resonant to DXs. The IX long-range transport vanishes at temperatures $\gtrsim 10$~K. With increasing IX density, IX localization, then IX long-range transport, and then IX reentrant localization is observed.  The results are in qualitative agreement with the Bose-Hubbard theory of bosons in periodic potentials predicting superfluid at $N \sim 1/2$ and insulating at $N \sim 0$ and $N \sim 1$ phases for the number of bosons per site of the periodic potential $N$.

\section{Acknowledgments}

We thank M.M.~Fogler and J.R.~Leonard for discussions. We especially thank A.H.~MacDonald for discussions of IXs in moir{\'e} potentials and A.K.~Geim for teaching us manufacturing TMD HS. The studies were supported by DOE Office of Basic Energy Sciences under Award DE-FG02-07ER46449. The HS manufacturing was supported by NSF Grant 1905478.

\vskip 5 mm
$^*$equal contribution

\end{document}


\date{July 1, 2023}

\title{Supporting Information for 
\\Transport and localization of indirect excitons in a van der Waals heterostructure
}

\author{L.~H.~Fowler-Gerace$^*$}
\affiliation{Department of Physics, University of California at San Diego, La Jolla, CA 92093, USA}
\author{Zhiwen~Zhou$^*$}
\affiliation{Department of Physics, University of California at San Diego, La Jolla, CA 92093, USA}
\author{E.~A.~Szwed}
\affiliation{Department of Physics, University of California at San Diego, La Jolla, CA 92093, USA}
\author{D.J. Choksy}
\affiliation{Department of Physics, University of California at San Diego, La Jolla, CA 92093, USA}
\author{L.~V.~Butov} 
\affiliation{Department of Physics, University of California at San Diego, La Jolla, CA 92093, USA}

\begin{abstract}
\noindent
\end{abstract}

\maketitle

\renewcommand*{\thefigure}{S\arabic{figure}}

\subsection{Supplementary Notes 1: The heterostructure details}

The van der Waals heterostructure was assembled using the dry-transfer peel-and-lift technique~\cite{Withers2015}. Crystals of hBN, MoSe$_2$, and WSe$_2$ were first mechanically exfoliated onto different Si substrates that were coated with a double polymer layer consisting of polymethyl glutarimide (PMGI) and polymethyl methacrylate (PMMA). The bottom PMGI was then dissolved with the tetramethylammonium hydroxide based solvent CD-26, causing the top PMMA membrane with the target 2D crystal to float on top of the solvent. The PMMA membrane functions both as a support substrate for transferring the crystal and as a barrier to protect the crystal from the solvent. Separately, a large graphite crystal was exfoliated onto an oxidized Si wafer, which served as the basis for the heterostructure. The PMMA membrane supporting the target crystal was then flipped over and aligned above a flat region of the graphite crystal using a micromechanical transfer stage. The two crystals were brought into contact and the temperature of the stage was ramped to $80^{\circ}$~C in order to increase adhesion between the 2D crystals. Then, the PMMA membrane was peeled off leaving the bilayer stack on the wafer. The procedure was repeated leading to a multicrystal stack with the desired layer sequence. Sample annealing was performed by immersing the sample in Remover PG, an N-methyl-2-pyrrolidone (NMP) based solvent stripper, at $70^{\circ}$~C for 12 hours.

\begin{figure}
\begin{center}
\includegraphics[width=13.5cm]{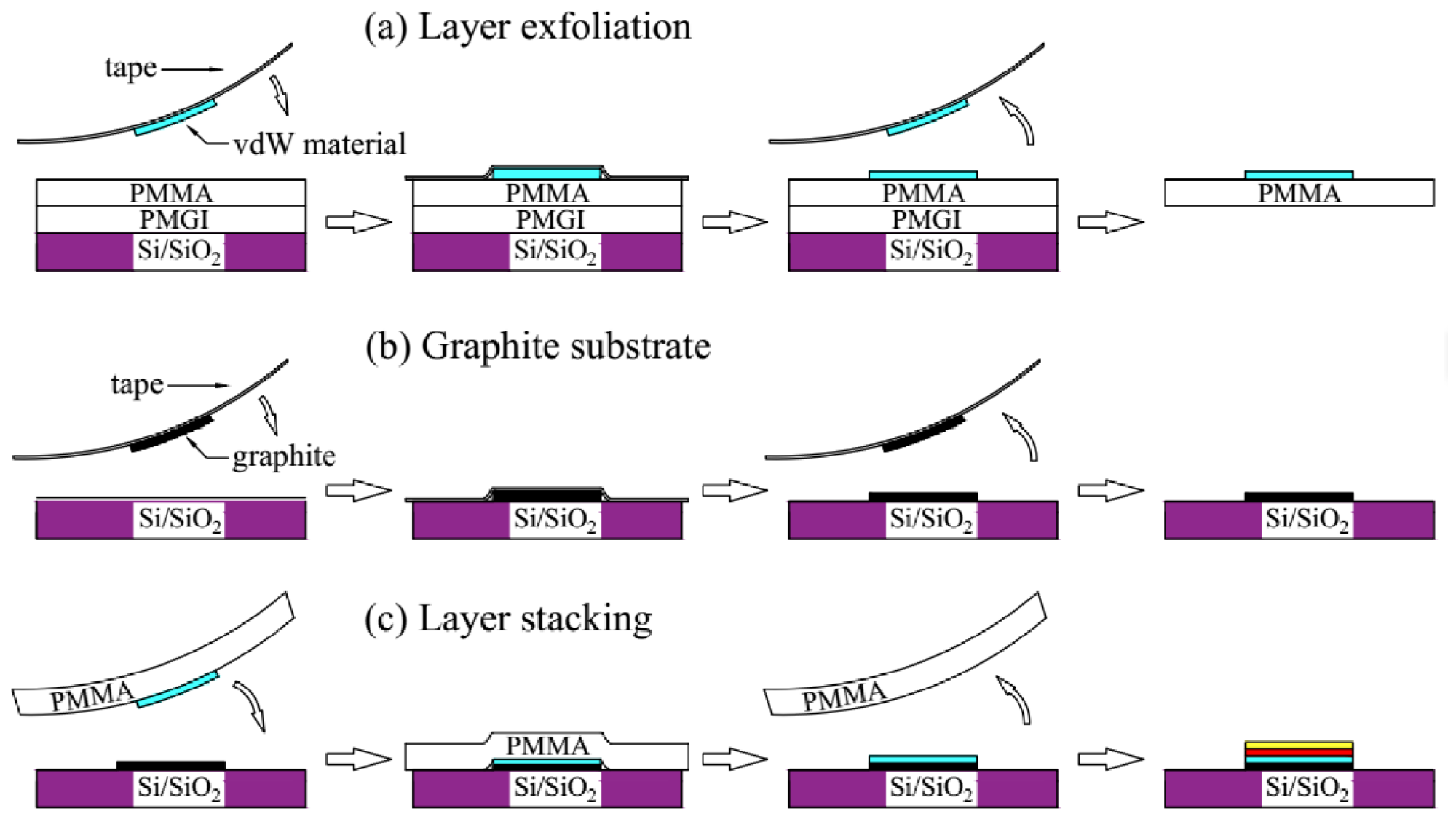}
\caption{Sample fabrication process. (a) Van der Waals layer exfoliation. (b) Graphite substrate exfoliation. (c) Layer stacking.
}
\end{center}
\label{fig:spectra}
\end{figure}

No intentional sample doping was done; however, unintentional $n-$type doping is typical for TMD layers~\cite{Withers2015}. The thickness of bottom and top hBN layers is about 40 and 30~nm, respectively. The MoSe$_2$ layer is on top of the WSe$_2$ layer. The long WSe$_2$ and MoSe$_2$ edges reach $\sim 30$ and $\sim 20$~$\mu$m, respectively, which enables a rotational alignment between the WSe$_2$ and MoSe$_2$ monolayers. 
The twist angle $\delta \theta = 1.1^\circ$ corresponding to the moir{\'e} superlattice period $b = 17$~nm, which gives $N \sim 1/2$ at the estimated $n \sim 2 \times 10^{11}$~cm$^{-2}$ for the long-range IX transport as outlined in the main text, agrees with the angle between MoSe$_2$ and WSe$_2$ edges in the HS (Fig.~S2c). 

The accuracies of estimating $\delta \theta$ using the long WSe$_2$ and MoSe$_2$ edges and using SHG are comparable. We do not use SHG for additional estimates of $\delta \theta$ since the intense optical excitation pulses in SHG measurements may cause a deterioration of the HS and may suppress the long-range IX transport. We note that $\delta \theta$ and, in turn, the moir{\'e} potential may vary over the HS area.

Figure~S2 presents a microscope image showing the layer pattern of the HS. The layer boundaries are indicated. The hBN layers cover the entire areas of MoSe$_2$ and WSe$_2$ layers. There was a narrow multilayer graphene electrode on the top of the HS around $x = 2$~$\mu$m for $y=0$, Fig.~S2. This electrode was detached. The IX luminescence reduction around $x =2$~$\mu$m can be related with residual graphene layers on the HS.

\begin{figure}
\begin{center}
\includegraphics[width=15cm]{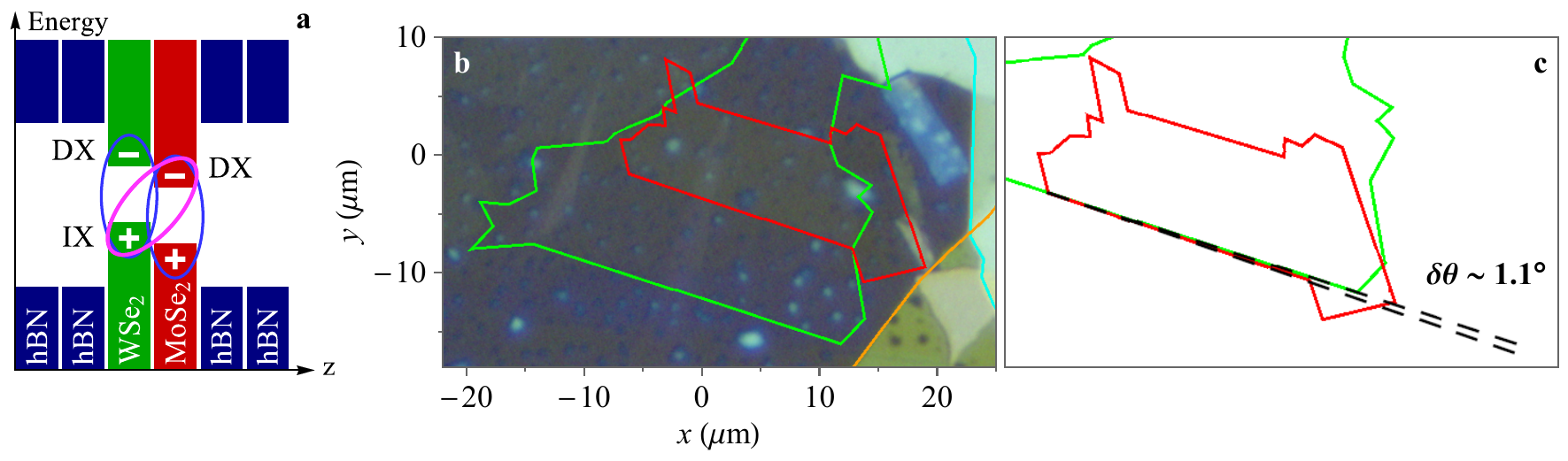}
\caption{(a) Schematic energy-band diagram for the HS. The ovals indicate a direct exciton (DX) and an indirect exciton (IX) composed of an electron (--) and a hole (+).
(b) A microscope image showing the layer pattern of the HS. The green, red, cyan, and orange lines indicate the boundaries of WSe$_2$ and MoSe$_2$ monolayers and bottom and top hBN layers, respectively. 
(c) A schematic showing the angle between MoSe$_2$ and WSe$_2$ edges in the HS. The layers contours are shifted keeping their orientation. The angle between MoSe$_2$ and WSe$_2$ edges in the HS is compared with $\delta \theta = 1.1^\circ$ (the angle between the dashed lines).
}
\end{center}
\label{fig:spectra}
\end{figure}

The discussions of IX transport and localization in this work apply to both R and H (AA and AB) stacking: the predicted amplitude of moir{\'e} superlattice potential for MoSe$_2$/WSe$_2$ HS with R and H stacking is $\sim 100$ and $\sim 25$~meV, respectively~\cite{Yu2017, Wu2018} and both these values are much larger than the interaction energy $\delta E \sim 3$~meV and thermal energy $k_{\rm B}T \lesssim 1$~meV for the observed long-range IX transport. We did not verify if stacking in the sample is R or H. As outlined in the main text, the observed IX transport phenomena are qualitatively similar for R and H stacking. It is important to verify the role of the rotational alignment between the HS layers on the IX localization and transport. So far the long-range IX transport with the decay distance $d_{1/e}$ exceeding 100~$\mu$m was realized in one sample in this work. Other samples show $d_{1/e}$ up to a few $\mu$m~\cite{Rivera2016, Jauregui2019, Unuchek2019a, Unuchek2019, Liu2019, Choi2020, Huang2020, Yuan2020, Li2021, Wang2021, Shanks2022, Sun2022, Tagarelli2023}.
The shorter range of IX transport likely originates from disorder, which suppresses conducting phases~\cite{Fisher1989}. 
The sample studied in this work demonstrates the existence of the IX long-range transport and reentrant localization and allows measuring the density-temperature phase diagram for these phenomena and comparing it with the theory. However, it is essential to study these phenomena in other samples and, in particular, to study the dependence on the twist angle. The sample statistics outlined above show that it is challenging to manufacture samples with different twist angles, all with sufficiently small disorder, in order to study the dependence of the IX localization and transport on the twist angle and, in turn, on the moir{\'e} superlattice period. Consequently, this remains the subject for future works.

\subsection{Supplementary Notes 2: Optical measurements}

In the cw experiments, excitons were generated by a cw Ti:Sapphire laser with tunable excitation energy or a cw HeNe laser with excitation energy $E_{\rm ex} = 1.96$~eV. Luminescence spectra were measured using a spectrometer with resolution 0.2~meV and a liquid-nitrogen-cooled CCD. The spatial profiles of IX luminescence vs. $x$ were obtained from the luminescence images detected using the CCD. The signal was integrated from $y = - 0.5$ to $y = + 0.5$~$\mu$m. Figure S3 shows a representative luminescence image. Figure S4 shows spectra for resonant and non-resonant excitation. 

\begin{figure}
\begin{center}
\includegraphics[width=5.5cm]{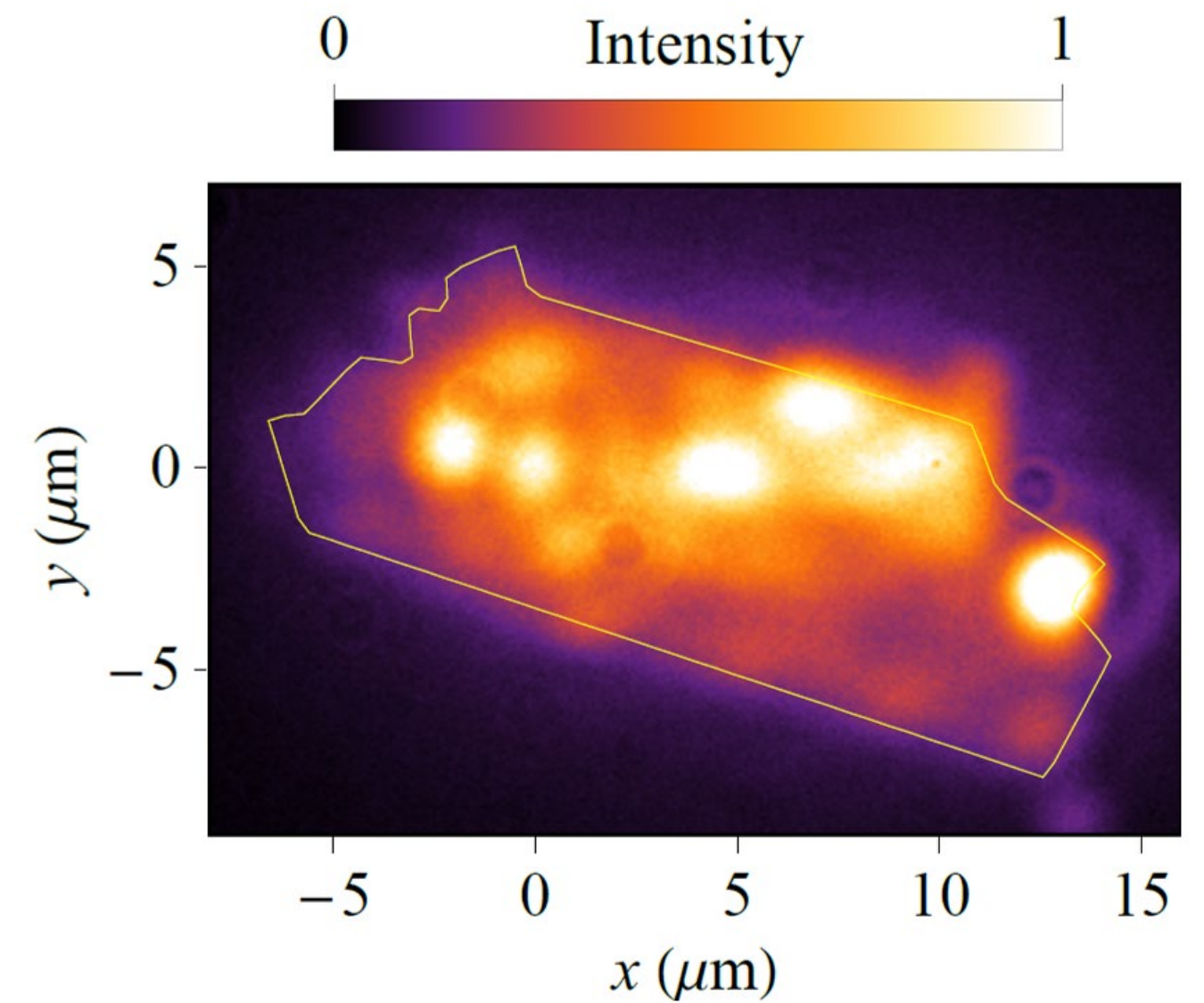}
\caption{IX luminescence image at $P_{\rm ex} = 0.2$~mW, $T = 6$~K, $E_{\rm ex} = 1.623$~eV. The $\sim$ 1.5 $\mu$m laser spot is centered at $x = 0$, $y = 0$.
}
\end{center}
\label{fig:spectra}
\end{figure}

\begin{figure}
\begin{center}
\includegraphics[width=5.5cm]{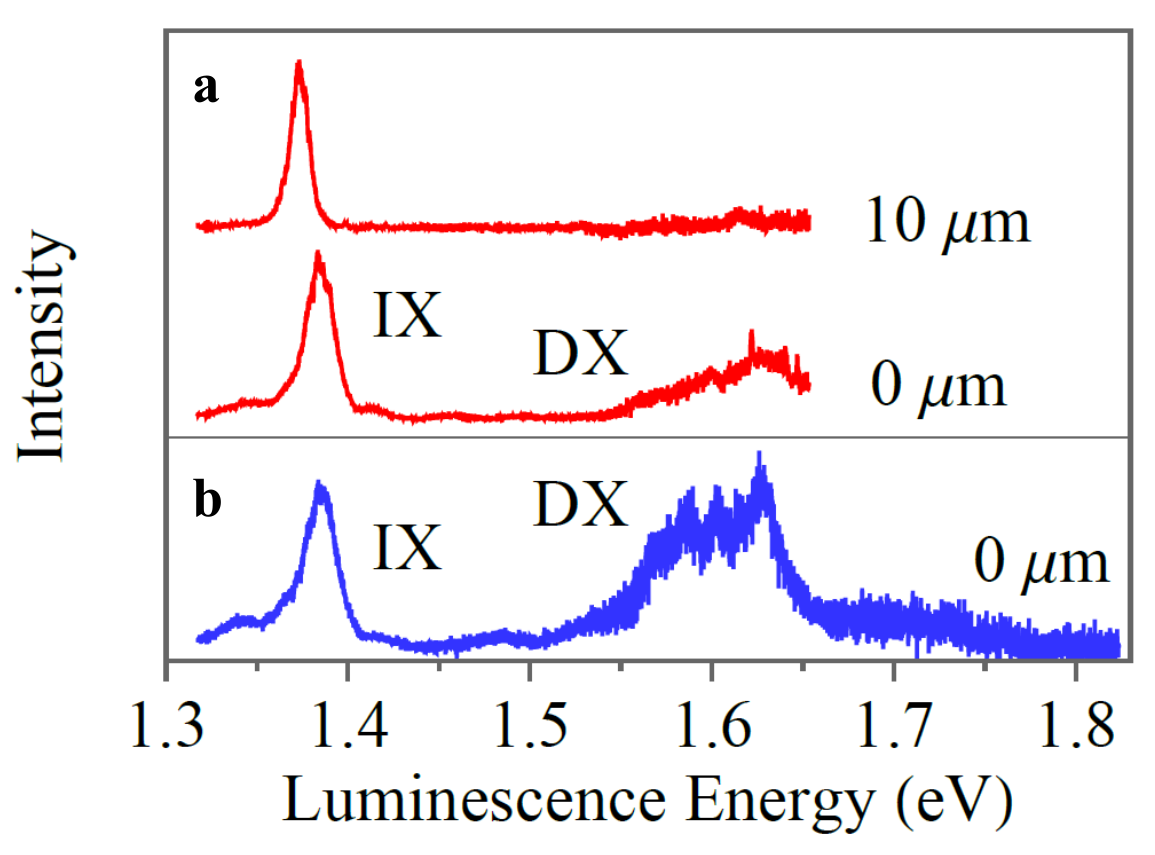}
\caption{The spectra for resonant [$E_{\rm ex} = 1.694$~eV, (a)] and non-resonant [$E_{\rm ex} = 1.96$~eV, (b)] excitation. The relative intensity of the higher-energy DX luminescence is lower for the resonant excitation, evidencing a lower temperature of the exciton system. $P_{\rm ex} = 0.2$~mW. $T = 1.7$~K. The $\sim$ 1.5 $\mu$m laser spot is centered at $x = 0$, $y = 0$.
}
\end{center}
\label{fig:spectra}
\end{figure}

The IX luminescence kinetics was measured using a pulsed semiconductor laser with $E_{\rm ex} = 1.694$~eV nearly resonant to WSe$_2$ DX energy. The emitted light was detected by a liquid-nitrogen-cooled CCD coupled to a PicoStar HR TauTec time-gated intensifier. 

The experiments were performed in a variable-temperature 4He cryostat. The sample was mounted on an Attocube xyz piezo translation stage allowing adjusting the sample position relative to a focusing lens inside the cryostat. 
All physical phenomena reported in the paper are reproducible after ca.~100 cooling down to 2~K and warming up to room temperature.

\subsection{Supplementary Notes 3: The drift-diffusion model of IX transport}

This section outlines the drift-diffusion model describing classical diffusive IX transport~\cite{Ivanov2002, Ivanov2006, Hammack2009}.
Within this model, IX transport is described by the equation for IX density $n$
\begin{eqnarray}
\frac{\partial n}{\partial t} = \nabla \left[D\nabla n + \mu n \nabla (u_0 n)\right] + \Lambda - \frac{n}{\tau}\,
\end{eqnarray}
The first and second terms in square brackets in Eq.~1 describe IX diffusion and drift currents, respectively. The latter originates from the IX repulsive dipolar interactions and is approximated by the mean-field 'plate capacitor' formula for the IX energy shift with density $\delta E = n u_0$, $u_0 = 4 \pi e^2 d_z/\varepsilon$~\cite{Yoshioka1990}. The diffusion coefficient 
\begin{eqnarray}
D = D^{(0)} \exp [ - U^{(0)}/(k_{\rm B}T + n u_0)] 
\end{eqnarray}
accounts for the temperature- and density-dependent screening of the long-range-correlated in-plane potential landscape by interacting IXs~\cite{Ivanov2002, Ivanov2006, Hammack2009}. $D^{(0)}$ is the diffusion coefficient in the absence of in-plane potential. $U^{(0)}/2$ is the amplitude of the in-plane potential. 
$\mu$ is the IX mobility. The IX generation rate $\Lambda$ has a profile of the laser excitation spot. $\tau$ is the IX lifetime. 

\begin{figure}
\begin{center}
\includegraphics[width=7.5cm]{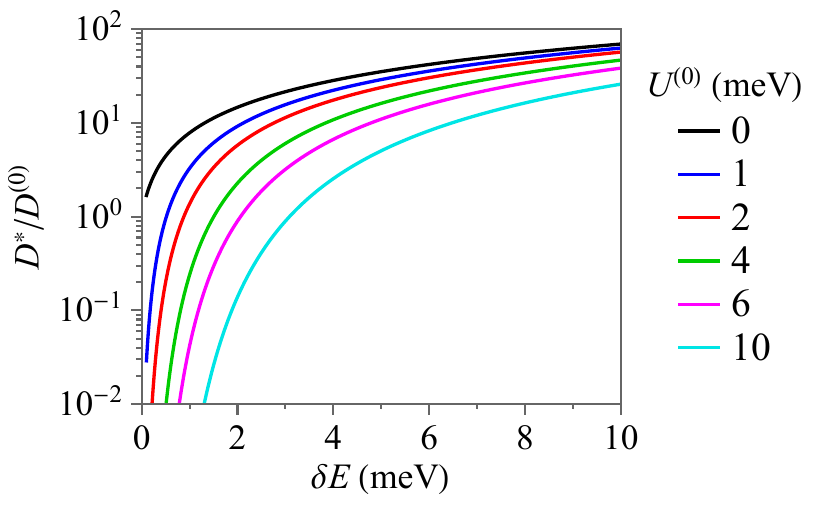}
\caption{$D^*/D^{(0)}$ vs. $\delta E$ for different $U^{(0)}$. $T = 1.7$~K.
}
\end{center}
\label{fig:spectra}
\end{figure}

\begin{figure}
\begin{center}
\includegraphics[width=15cm]{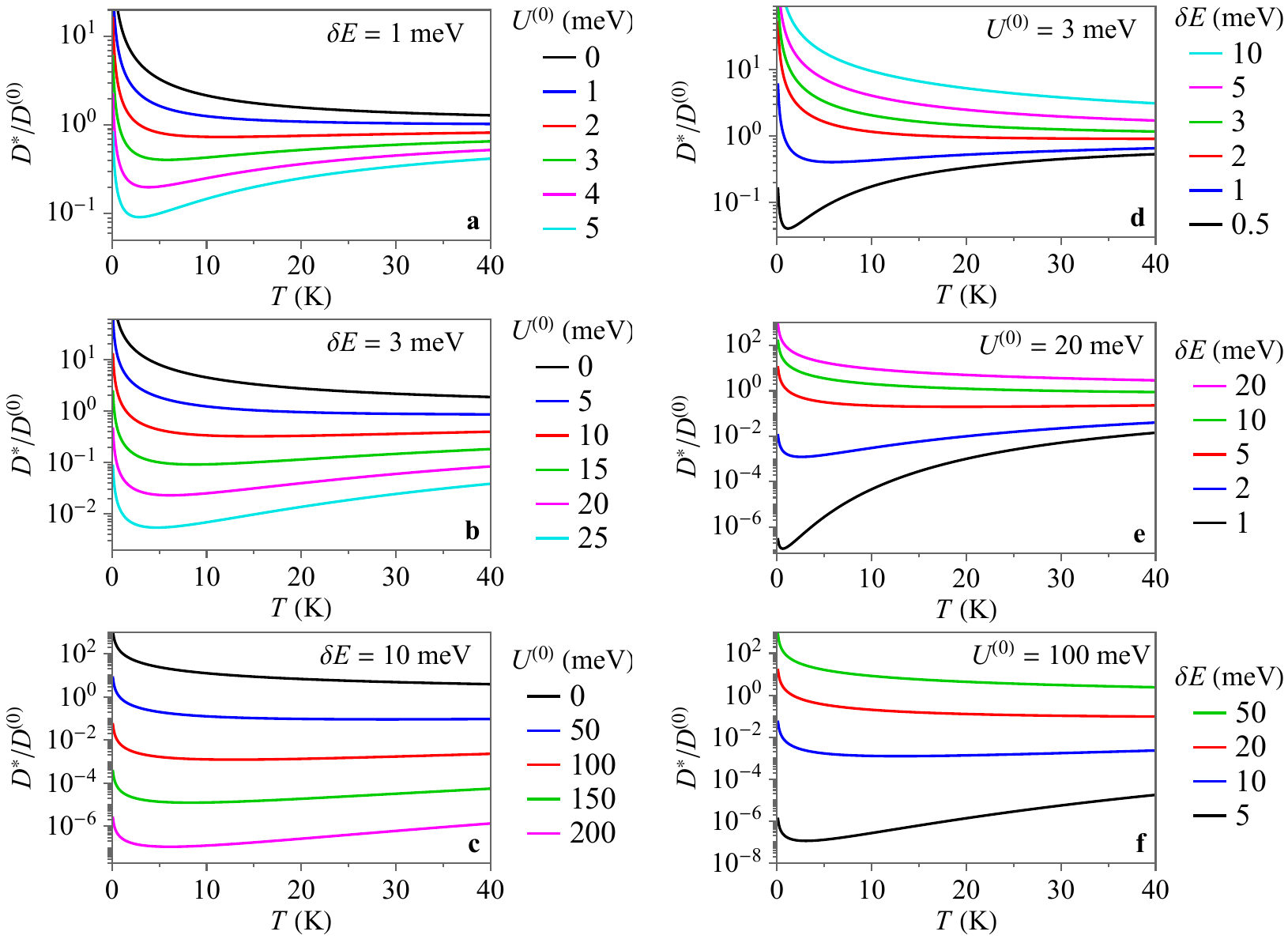}
\caption{$D^*/D^{(0)}$ vs. $T$ for different $\delta E$ and $U^{(0)}$.
}
\end{center}
\label{fig:spectra}
\end{figure}

Both the IX-interaction-induced screening of in-plane potential and the IX-interaction-induced drift from the origin contribute to an enhancement of IX transport with increasing IX density $n$. In particular, within the classical drift-diffusion model~\cite{Ivanov2002, Ivanov2006, Hammack2009}, the enhancement of IX transport due to the IX-interaction-induced screening of in-plane potential is described by Eq.~2. Fitting the IX cloud expansion by $R^2 \sim D^* \delta t$ probes the effective IX diffusion coefficient $D^* = D + \mu n u_0$, which includes both the diffusion and the drift due to the density gradient~\cite{Dorow2017}. 
$\mu$ can be estimated using the Einstein relation $\mu = D / (k_BT)$. Therefore, the enhancement of IX transport due to the IX-interaction-induced drift from the origin is described by the enhanced $D^*$ (Eq.~3)~\cite{Dorow2017}
\begin{eqnarray}
D^* = D [1 + n u_0/ (k_BT)]
\end{eqnarray}
Equations~2 and 3 show that within the classical drift-diffusion model~\cite{Ivanov2002, Ivanov2006, Hammack2009}, the IX transport should enhance with density for any amplitude of the in-plane potential and for any temperature. Figure~S5 shows an enhancement of $D^*$ with density for different $U^{(0)}$. The classical IX drift and diffusion~\cite{Ivanov2002, Ivanov2006, Hammack2009} (Eqs.~1-3, Fig.~S5) is inconsistent with the observed reduction of IX transport with density at high densities (Fig.~3a).

Equations~2 and 3 show that within the classical drift-diffusion model~\cite{Ivanov2002, Ivanov2006, Hammack2009}, $D$ should increase with temperature for any $U^{(0)}$ and the dependence of $D^*$ on temperature is nonmonotonic (Fig.~S6). Within this model, $D^*$ increases with temperature when $U^{(0)} > \delta E [1 + \delta E/(k_{\rm B} T)]$. The classical IX drift and diffusion model~\cite{Ivanov2002, Ivanov2006, Hammack2009} (Eqs.~1-3, Fig.~S6) is inconsistent with the observed vanishing of IX transport with temperature (Fig.~3b). Within the classical drift-diffusion model~\cite{Ivanov2002, Ivanov2006, Hammack2009}, the IX transport should enhance with the temperature for the strong moir{\'e} potentials~\cite{Yu2017, Wu2018}. 

The drift-diffusion Eq.~1 can be supplemented by the thermalization equation, which describes heating of excitons by photoexcitation and cooling via interaction with phonons, and by including the HS in-plane potential in the drift term~\cite{Ivanov2006, Hammack2009}. In particular, the exciton thermalization can lead to the appearance of the inner ring in exciton luminescence patterns~\cite{Ivanov2006, Hammack2009}. The Einstein relation can be extended to the generalized Einstein relation $\mu = D (e^{T_{\rm q}/T} - 1)/(k_{\rm B}T_{\rm q})$, which gives $\mu = D/(k_BT)$ for the classical drift and diffusion~\cite{Ivanov2002, Ivanov2006, Hammack2009}.

\subsection{Supplementary Notes 4: IX interaction energies and the long-range IX transport}

Below, we briefly discuss the IX interaction energies $\delta E$ corresponding to the long-range IX transport. In GaAs HS, the strong enhancement of IX transport, the IX delocalization, is observed when the IX interaction energy becomes comparable to the amplitude of the in-plane potential, which is, in turn, comparable to the IX luminescence linewidth at low IX densities~\cite{Remeika2009}. For the long-range IX transport in the MoSe$_2$/WSe$_2$ HS, the IX interaction energy $\delta E \sim 3$~meV (Fig.~2b). This value is comparable to the smallest IX linewidth $\sim 4$~meV at low IX densities in the MoSe$_2$/WSe$_2$ HS (Fig.~S7a). However, the IX interaction energy $\delta E \sim 3$~meV is significantly smaller than the predicted IX energy modulations in moir{\'e} superlattice potentials in MoSe$_2$/WSe$_2$ HS that are in the range of tens of meV~\cite{Wu2018, Yu2017}. This also shows that the IX transport in MoSe$_2$/WSe$_2$ HS with the moir{\'e} superlattice potentials is distinct from the IX transport in GaAs HS. The classical drift and diffusion model~\cite{Remeika2009} does not describe the quantum transport of repulsively interacting particles in periodic potentials. Sharp emission lines appear in the IX spectrum at lower densities (Fig.~S7b). The sharp lines will be considered elsewhere.

\begin{figure}
\begin{center}
\includegraphics[width=10cm]{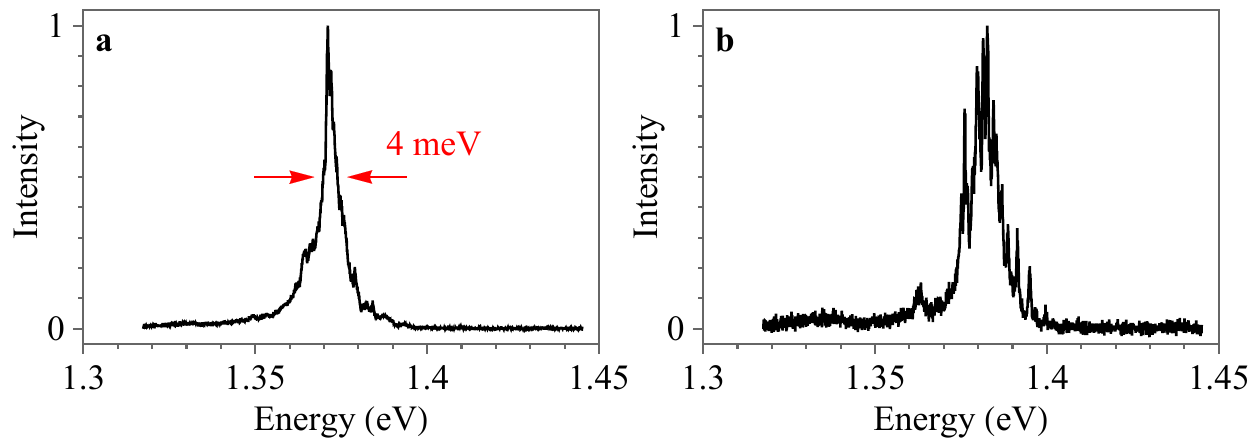}
\caption{(a) IX luminescence spectrum at a low IX density in the MoSe$_2$/WSe$_2$ HS. $P_{\rm ex} = 5$~$\mu$W, $T = 1.7$~K, $E_{\rm ex} = 1.96$~eV. (b) IX luminescence spectrum showing sharp emission lines at lower IX densities in the MoSe$_2$/WSe$_2$ HS. $P_{\rm ex} = 1$~$\mu$W, $T = 1.7$~K, $E_{\rm ex} = 1.689$~eV.
}
\end{center}
\label{fig:spectra}
\end{figure}

\begin{figure}
\begin{center}
\includegraphics[width=5.5cm]{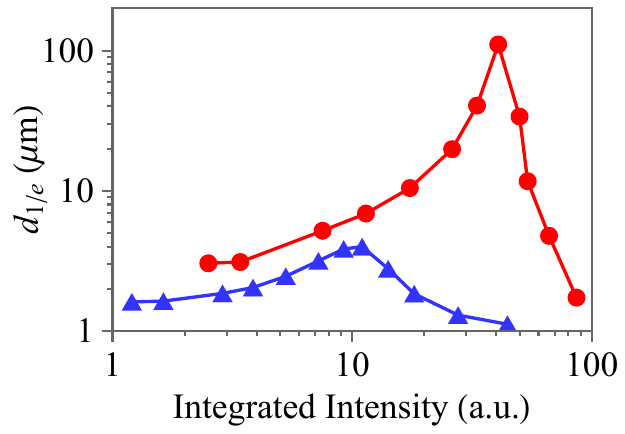}
\caption{$d_{1/e}$ vs. integrated IX intensity in the entire MoSe$_2$/WSe$_2$ HS. Excitation is either nonresonant ($E_{\rm ex} = 1.96$~eV, blue), or near resonant ($E_{\rm ex} = 1.623$~eV, red) to the DX. $T = 6$~K.
}
\end{center}
\label{fig:spectra}
\end{figure}

\subsection{Supplementary Notes 5: IX transport decay distance for nearly resonant excitation and nonresonant excitation}

The nearly resonant excitation produces a higher IX density and, in turn, a stronger IX luminescence signal due to a higher absorption (Fig.~1). The higher IX densities can be also achieved by nonresonant excitation with higher excitation powers $P_{\rm ex}$. 
Figure~2b shows that for the same $\delta E$, a much higher $d_{1/e}$ is realized for the resonant excitation and that the strong enhancement of IX propagation at resonant excitation originates from the suppression of localization and scattering of IXs, as described in the main text. 
This is confirmed by Fig.~S8, which also compares the IX transport decay distances $d_{1/e}$ for the nearly resonant and nonresonant excitations. Figure~S8 shows that in the regime of the long-range IX transport at the nearly resonant excitation, for the same IX signal in the HS, achieved by a higher $P_{\rm ex}$ at nonresonant excitation, $d_{1/e}$ is significantly longer in the case of the nearly resonant excitation. This indicates that the nearly resonant excitation is essential for the achievement of the long-range IX transport.

\subsection{Supplementary Notes 6: IX decay kinetics}

Figure~S9 shows the IX luminescence decay kinetics after the laser excitation pulse is off. The IX decay times $\tau$ (Fig.~S9) are orders of magnitude longer than the DX decay times~\cite{Korn2011}. Both increasing $P_{\rm ex}$ (Fig.~S9a,c) and temperature (Fig.~S9b,d) lead to a reduction of $\tau$. This variation of $\tau$ is rather weak and its effect on the IX transport variation is minor. For instance, $\tau$ monotonically varies with $P_{\rm ex}$ (Fig.~S9c) while the IX transport distance first increases and then reduces with $P_{\rm ex}$ (Fig.~3a) and the variation of $\tau$ (Fig.~S9c,d) is significantly smaller than the variation of $d_{1/e}$ (Fig.~3).

\begin{figure}
\begin{center}
\includegraphics[width=8cm]{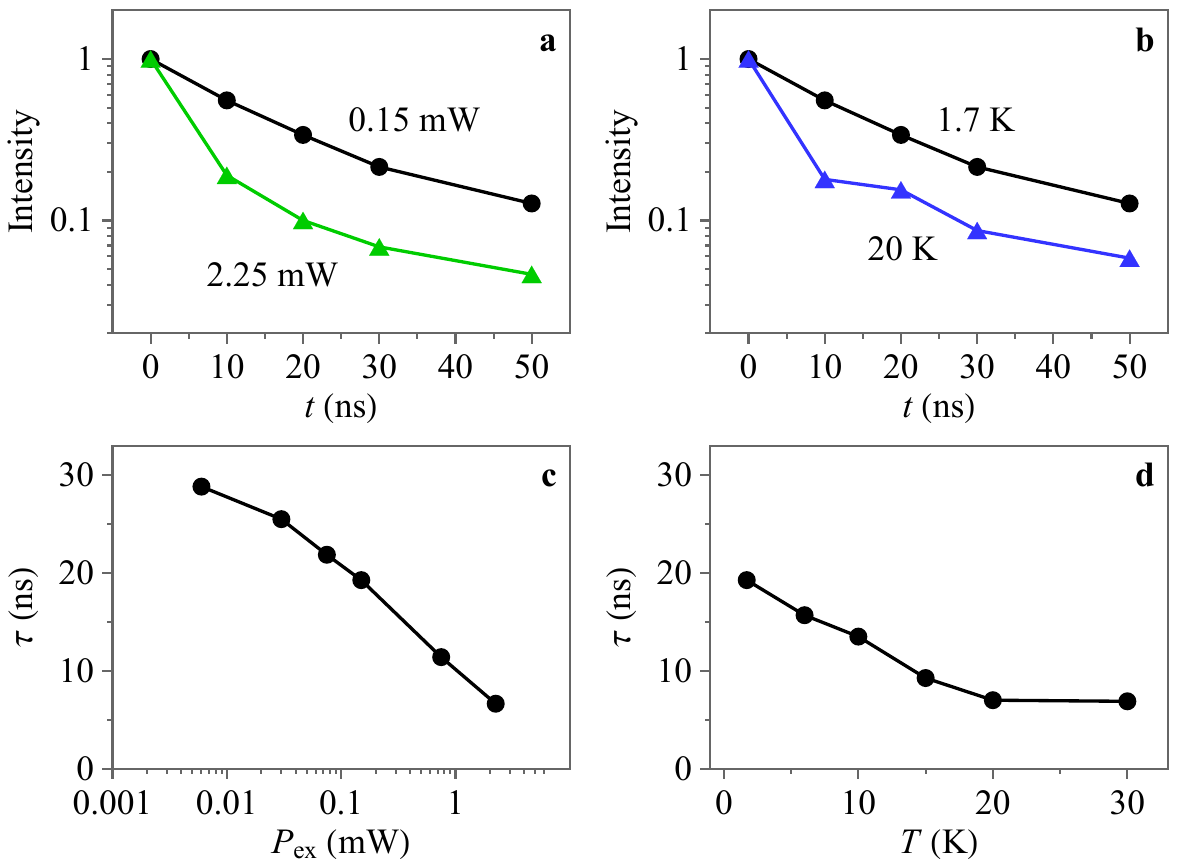}
\caption{IX decay kinetics. (a,b) Spatially integrated IX luminescence intensity vs. time for different $P_{\rm ex}$ (a) and temperatures (b). The times are given at the ends of the 10 ns signal integration windows. The excitation pulse ends at $t=0$. (c,d) The IX lifetime $\tau$ vs. $P_{\rm ex}$ (c) and vs. temperature (d). $\tau$ is the initial decay time after the excitation pulse end. $T = 1.7$~K (a,c), $P_{\rm ex} = 0.15$~mW (b,d). For all data, $E_{\rm ex} = 1.694$~eV. 
}
\end{center}
\label{fig:spectra}
\end{figure}

\subsection{Supplementary Notes 7: Rough estimates of the radiative and nonradiative lifetime variation with $P_{\rm ex}$ and temperature}

The IX radiative and nonradiative lifetimes, $\tau_{\rm r}$ and $\tau_{\rm nr}$, can be roughly estimated from the measured IX total luminescence intensity $I$ and decay time $\tau$. Figure~S10 shows $\tau_{\rm r}$ and $\tau_{\rm nr}$ estimated using $\tau_{\rm r}^{-1} = (I / \Lambda) \tau^{-1}$ and $\tau_{\rm nr}^{-1} = \tau^{-1} - \tau_{\rm r}^{-1}$. Within this estimate, the IX luminescence kinetics is approximated as monoexponential. It is also assumed that there is no IX spatial escape from the system. This is approximated by integrating the IX luminescence signal over the heterostructure. For an estimate of the generation rate $\Lambda$, we use $I_{\rm D} = \alpha \Lambda$, where $I_{\rm D}$ is the DX luminescence signal for the excitation at the MoSe$_2$ monolayer outside the heterostructure and $\alpha$ is the quantum efficiency for this DX luminescence. Qualitatively similar variations of $\tau_{\rm r}$ and $\tau_{\rm nr}$ with $P_{\rm ex}$ and temperature are obtained for different values of $\alpha$ (Fig.~S10). 

\begin{figure}
\begin{center}
\includegraphics[width=9cm]{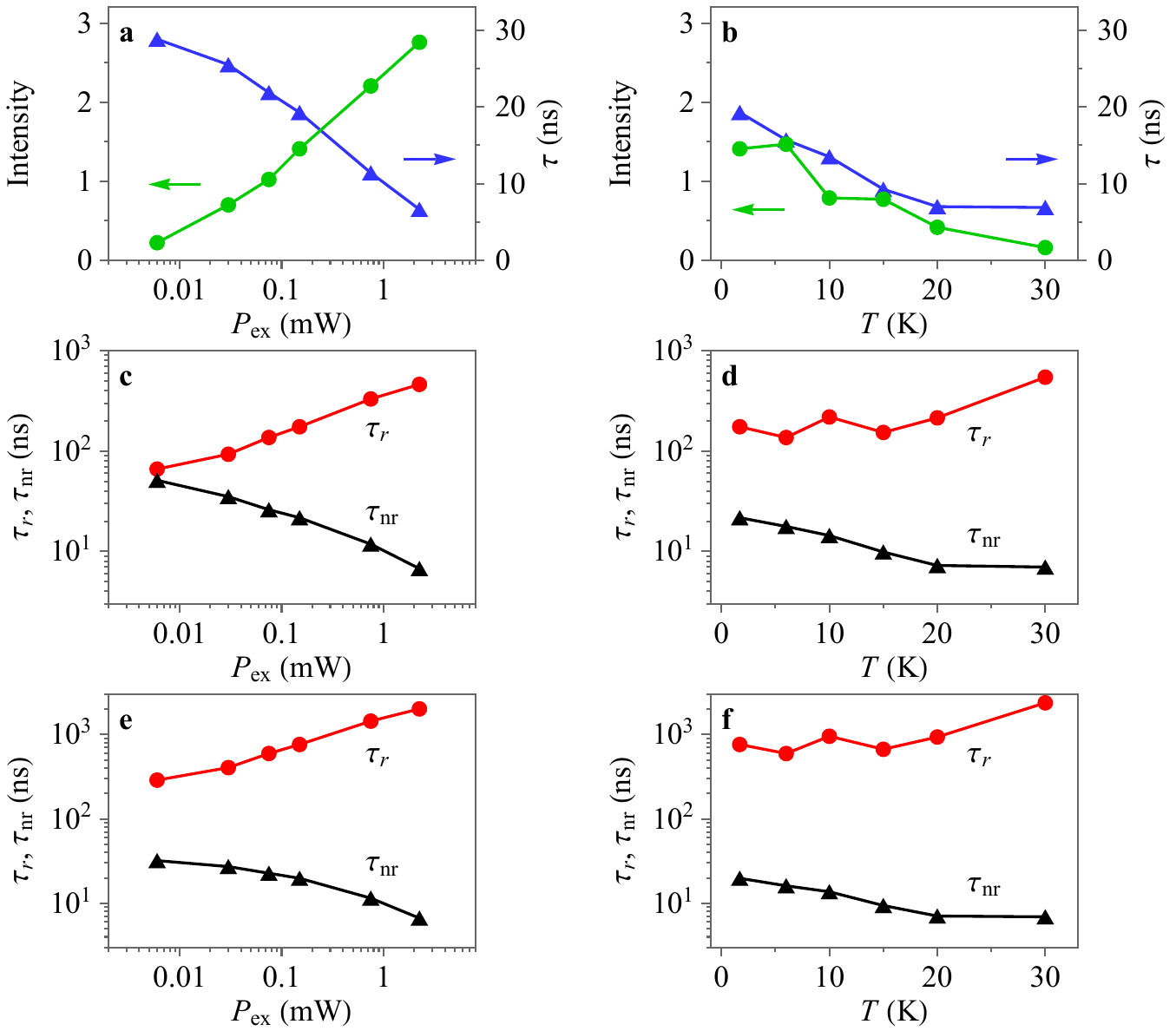}
\caption{Rough estimate of radiative and nonradiative lifetimes, $\tau_{\rm r}$ and $\tau_{\rm nr}$. (a,b) Spatially integrated IX luminescence intensity $I$ (green) and IX lifetime $\tau$ (blue) vs. $P_{\rm ex}$ (a) and temperature (b). $T = 1.7$~K (a), $P_{\rm ex} = 0.15$~mW (b), and $E_{\rm ex} = 1.694$~eV (a,b). $\tau$ is the initial decay time after the excitation pulse end. (c-f) Estimated $\tau_{\rm r}$ and $\tau_{\rm nr}$ vs. $P_{\rm ex}$ (c,e) and temperature (d,f) for the quantum efficiency of DX luminescence in the MoSe$_2$ monolayer $\alpha = 100$\% (c,d) and 25\% (e,f).
}
\end{center}
\label{fig:spectra}
\end{figure}

\subsection{Supplementary Notes 8: IX transport kinetics}

\begin{figure}
\begin{center}
\includegraphics[width=8.5cm]{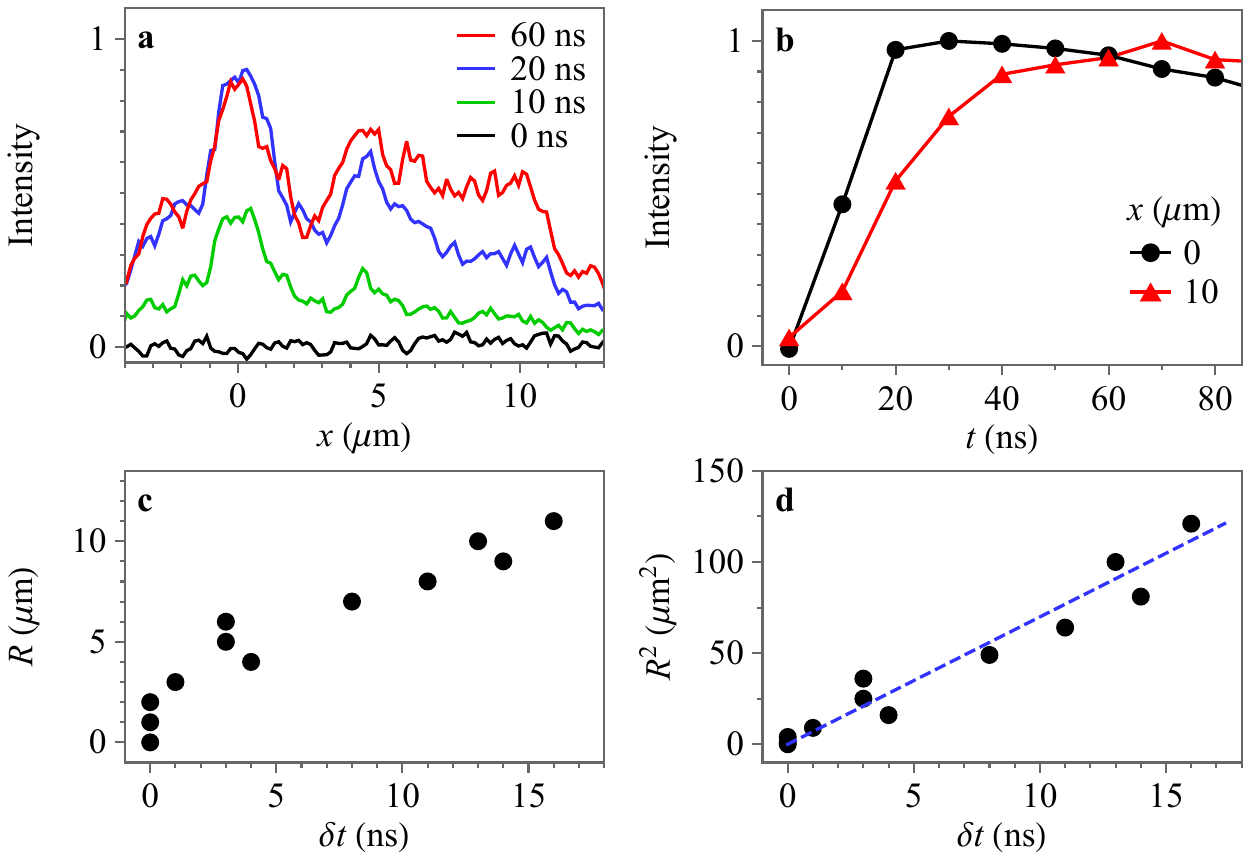}
\caption{IX transport kinetics. (a) IX luminescence profiles during the laser excitation pulse at different times. (b) Normalized IX luminescence intensity vs. time at different positions. (c) $R$ vs. $\delta t$. $\delta t$ is the time to reach 70\% of the maximum intensity at distance $R$ from the excitation spot relative to that time in the excitation spot. (d) $R^2$ vs. $\delta t$. The dashed line is a fit to the data with $D^* \sim R^2/\delta t \sim 70$~cm$^2$/s. $P_{\rm ex} = 0.15$~mW, $T = 1.7$~K, the times are given at the ends of the 10~ns signal integration windows, the excitation pulse starts at $t=0$, $E_{\rm ex} = 1.694$ eV, the $\sim 2$~$\mu$m laser spot is centered at $x = 0$.
}
\end{center}
\label{fig:spectra}
\end{figure}

Figure~S11 shows the kinetics of IX propagation from the excitation spot in the regime of long-range transport. The kinetics is measured during the rectangular-shaped laser excitation pulses with the duration 100~ns and period 300~ns. The 200~ns off time exceeds the IX lifetime (Fig.~S9) and is sufficient for a substantial decay of the IX signal. The IX luminescence at locations away from the excitation spot is delayed in comparison to the IX luminescence in the excitation spot (Fig.~S11a,b). The delay times $\delta t$ for the IX cloud to expand to the locations separated by distance $R$ from the origin allow estimating the IX transport characteristics. The substantial scattering of experimental points (Fig.~S11c,d) does not allow distinguishing transport with nearly constant velocity $R \propto \delta t$ from diffusive transport with $R^2 \propto \delta t$, therefore both approaches are probed. For the former approach, the estimated average velocity of the IX cloud expansion for the time range $\Delta t = 1 - 20$ ns, $v = \Delta R / \Delta t \sim 5 \times 10^4$~cm/s (Fig.~S11c). For the latter approach, fitting the IX cloud expansion by $R^2 \sim D^* \delta t$, gives $D^* \sim 70$~cm$^2$/s (Fig.~S11d). 
As described in Supplementary Notes 3, such fitting probes the effective IX diffusion coefficient $D^* = D + \mu n u_0$, which includes both the diffusion and the drift due to the density gradient~\cite{Dorow2017}. The IX mobility $\mu$ can be estimated using the Einstein relation $\mu = D / (k_BT)$, giving $D^* = D [1 + n u_0/(k_BT)]$. For $n u_0 \sim 3$~meV, $D^* \sim 70$~cm$^2$/s, and $T = 1.7$~K, this equation gives an estimate for the IX diffusion coefficient $D \sim 4$~cm$^2$/s and IX mobility $\mu = D / (k_BT) \sim 3 \times 10^4$~cm$^2$/(eV s). As outlined above, the observed IX transport is inconsistent with classical drift and diffusion, and these characteristics of IX transport are presented to allow a comparison with other studies of diffusive exciton transport.

\subsection{Supplementary Notes 9: IX luminescence profiles for different laser excitation energies}

Normalized IX luminescence profiles for different laser excitation energies $E_{\rm ex}$ are shown in Fig.~S12. The long-range IX transport with high decay distances $d_{1/e}$ is realized for $E_{\rm ex}$ close to the MoSe$_2$ or WSe$_2$ DX energy (Fig.~1). For some of the IX luminescence profiles no decay is observed within the HS. The data with the fit indicating $d_{1/e} > 100$~$\mu$m and showing no decay of the IX luminescence intensity with separation from the origin are presented in Fig.~1b by points on the edge.

\begin{figure}
\begin{center}
\includegraphics[width=8.5cm]{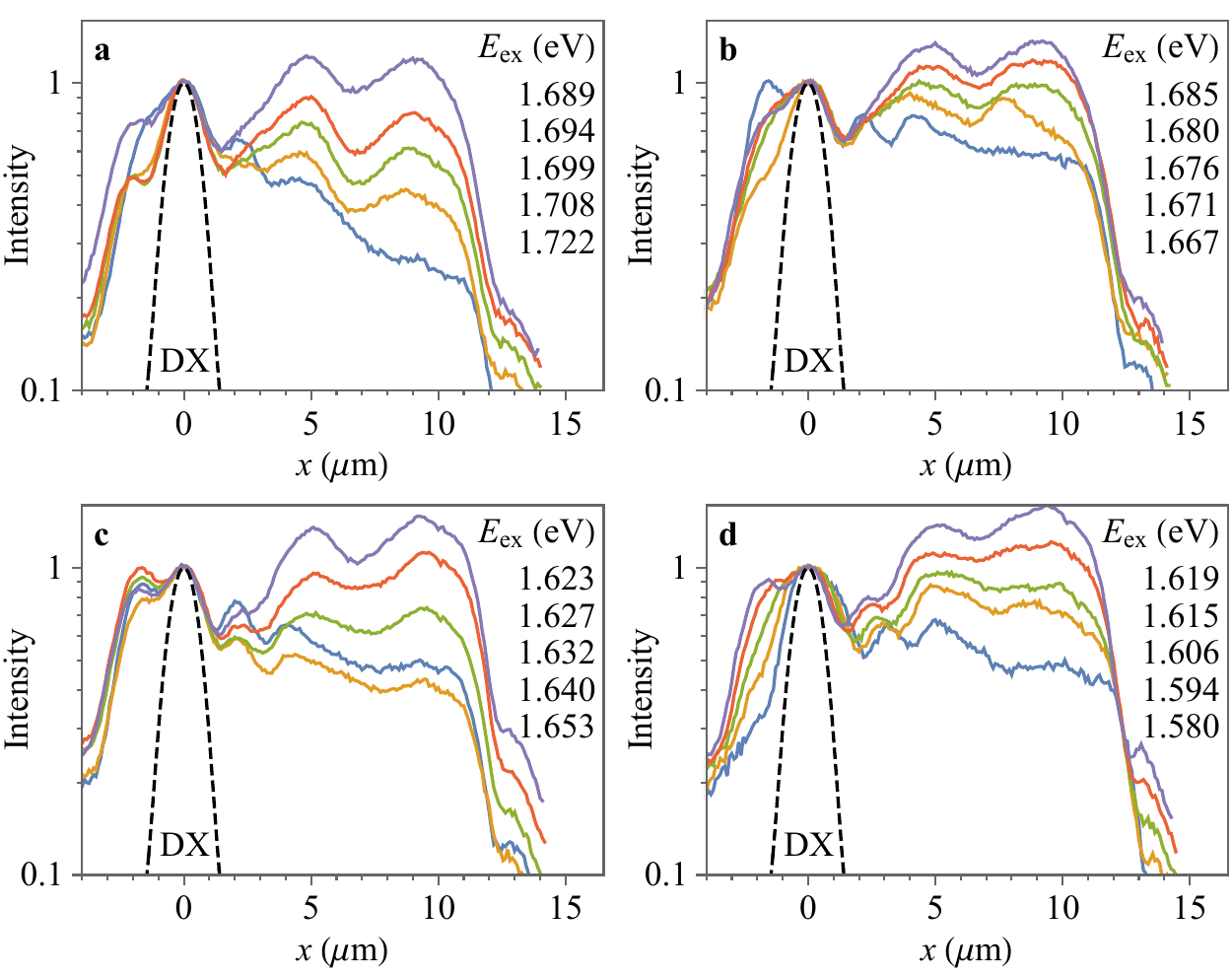}
\caption{Normalized IX luminescence profiles for different laser excitation energies $E_{\rm ex}$. The dashed line shows the DX luminescence profile in the WSe$_2$ monolayer, this profile is close to the laser excitation profile for a short DX propagation. $P_{\rm ex}$ = 0.2 mW, $T$ = 1.7 K, the $\sim$ 1.5 $\mu$m laser spot is centered at $x = 0$.
}
\end{center}
\label{fig:spectra}
\end{figure}

\vskip 5 mm
$^*$equal contribution